\documentstyle{laa}

\begin{document}

\title{The Sachs-Wolfe Effect}
 
\author{Martin White$^{1}$ and Wayne Hu$^{2}$}
\institute{$^{1}$Enrico Fermi Institute, University of Chicago,
5640 S. Ellis Ave., Chicago, IL 60637, USA\\
$^{2}$Institute for Advanced Study, Princeton NJ 08540, USA}

\thesaurus{12.03.1 12.03.4}
 
\date{Received 16 Sep 1996}
 
\maketitle\markboth{Martin White \& Wayne Hu: The Sachs-Wolfe Effect}{}
 
\begin{abstract}
We present a pedagogical derivation of the Sachs-Wolfe effect, specifically
the factor ${1\over 3}$ relating the temperature fluctuations to gravitational
potentials.  The result arises from a cancellation between gravitational
redshifts and intrinsic temperature fluctuations which can be derived from a
coordinate transformation of the background.
\end{abstract}

\section{Introduction}

On large scales, cosmic microwave background (CMB) anisotropies are related
to density fluctuations by the Sachs-Wolfe (\cite{SW}) effect.  
The gravitational effects of density perturbations on the potential $\Phi$
generates temperature fluctuations 
\begin{equation}
{\Delta T\over T} = -{1\over 3}\Phi \, ,
\label{eqn:simple}
\end{equation}
in the simplest case of adiabatic fluctuations in a matter dominated universe.

In this note we present an intuitive, but mathematically rigorous, derivation
of this formula and its generalizations.  While the effect is well known, a
simple but rigorous discussion does not appear to have been presented before.
The factor ${1\over 3}$ is easily derived from relativistic perturbation
theory (see e.g.~Mukhanov, Feldman \& Brandenberger \cite{MFB},
Liddle \& Lyth \cite{LidLyt}, Stebbins~\cite{Ste},
White, Scott \& Silk \cite{WSS}, Stoeger, Ellis \& Xu \cite{SEX},
Hu \cite{Hu}),
but that requires quite a high level of sophistication on the part of the
reader.  However these derivations exemplify the use of coordinate
transformations (Bardeen \cite{Bar}, Kodama \& Sasaki \cite{KodSas})
to expose the underlying physics.
Given the importance of the result for normalizing large-scale structure
models to the {\sl COBE} DMR measurement of large-angle anisotropies, we feel
a pedagogical introduction, using these ideas, is worthwhile.

\section{Derivation}

Following Sachs and Wolfe (\cite{SW}), we start from the geodesic equation
for photons propagating in a metric perturbed by a gravitational potential
$\Phi$. The resulting frequency shifts for the CMB photons lead to a
temperature perturbation
\begin{equation}
\left. {\Delta T\over T} \right|_f = \left. {\Delta T\over T}\right|_i
    - \Phi_i \, ,
\label{eqn:basic}
\end{equation}
where $i$ and $f$ refer to ``initial'' and ``final'' states.
We have dropped the term due to the local gravitational potential ($\Phi_f$)
which gives an isotropic temperature shift.
We also neglect the Doppler shift from the relative motion of the emitter and
receiver and other small scale effects, and assume that the potentials are
constant on large scales.  We return to these assumptions later.

The interpretation of Eq.~(\ref{eqn:basic}) is straightforward.  
The first term on the right-hand side is the ``intrinsic'' temperature
perturbation at early times.  The second term indicates the energy lost
when the photon climbs out of a potential well.
In this limit the Sachs-Wolfe effect is simply an expression of energy
conservation.

In order to rederive Eq.~(\ref{eqn:simple}), we consider the case of adiabatic
fluctuations in a critical density, matter dominated universe.
For adiabatic fluctuations an overdensity, or potential well, represents a
larger than average number of photons or an intrinsic hot spot.
We thus expect the two terms in Eq.~(\ref{eqn:basic}) to partially
cancel (see e.g.~Stebbins \cite{Ste}).
By comparison with Eq.~(\ref{eqn:simple}), we expect
$\Delta T/T|_i={2\over 3}\Phi$.

The derivation proceeds by moving to the rest frame of the cosmological fluid
(photons, baryons, dark matter \ldots) which is known as the ``comoving'' or
``velocity orthogonal isotropic'' gauge
(Bardeen \cite{Bar}, Kodama \& Sasaki \cite{KodSas}).
Here density fluctuations vanish and proper time coincides with coordinate
time at large scales.  The intrinsic term is negligible, which follows from
the Poisson equation: $\nabla^2\Phi=4\pi G\,\delta\rho$ or
$k^2\Phi=4\pi G\,a^2\,\delta\rho$ as $k\to0$.  Here $a(t)$ is the scale factor.
That proper time and coordinate time coincide follows from the fact that we
are in the rest frame of the fluid.
It is also in this frame that computation of fluctuations from inflation
takes on its simplest form
(Mukhanov et al.~\cite{MFB}, Liddle \& Lyth~\cite{LidLyt}).

However we wish to work in a frame where our Newtonian intuition makes sense,
the so called Newtonian gauge.  To get from our rest frame to the Newtonian
frame requires us to perform a shift of time coordinate.
(The spatial metric is isotropic in both the Newtonian frame and the rest
frame which does not allow us to make a redefinition of the space coordinates.)
Recall that in a gravitational potential clocks run slow
\begin{equation}
  ds = \sqrt{1-2\Phi}\ dt \simeq (1-\Phi) dt \, .
\label{eqn:slow}
\end{equation}
Since the background temperature is redshifting as $aT=\,$constant, in making
a shift of time coordinate $t\rightarrow t+dt$
(known also as a gauge transformation see Bardeen \cite{Bar} and
Kodama \& Sasaki \cite{KodSas}), we induce a temperature fluctuation.
If the equation of state is $p=w\rho$, then $a\sim t^{2/3(1+w)}$ and
\begin{equation}
\left. {\Delta T\over T}\right|_i =
  -{\delta a\over a} = {2\over 3(1+w)}\Phi \, ,
\label{eqn:intrinsic}
\end{equation}
where we have used Eq.~(\ref{eqn:slow}) in the last step.
For a matter dominated universe $w=0$, using Eq.~(\ref{eqn:basic}) gives
the factor of ${1\over 3}$ in Eq.~(\ref{eqn:simple}).  More generally,
\begin{equation}
\left. {\Delta T\over T}\right|_f = -{1 + 3w\over 3+3w}\Phi \, .
\end{equation}

There is another way to look at these two terms.  If the former discussion is
the ``fluid'' picture, this is the ``metric'' picture.  In this picture, the
second term in Eq.~(\ref{eqn:basic}) comes from the time-time part of the
metric.  Time dilation changes the frequency and hence the energy of
oscillators (see e.g.~Weinberg \cite{Wei}).  By contrast, the intrinsic term
is generated when we change from the rest frame to the Newtonian frame.
In so doing we change the definition of the spatial hypersurfaces and thus
the ``volume element'' or space-space part of the metric $\propto a^2$.
This induces a redshift just as the normal expansion of the universe induces
a redshift $\sim a^{-1}(t)$.  The change in the spatial curvature in changing
frames reproduces Eq.~(\ref{eqn:intrinsic}).

\section{Isocurvature}

Note that our argument works for isocurvature fluctuations also.
In this case, there are no initial curvature or metric perturbations, so
the frames coincide initially.  There are thus no initial perturbations in
the temperature at large scales.  Since the potentials do not vanish today,
our assumption of constant potentials needs to be relaxed.  This adds a
third term to Eq.~(\ref{eqn:basic}): if the potential changes while the photon
is crossing it, a net temperature shift remains between the infall blueshift
and the outclimb redshift.  A potential stretches space so the accompanying
changes in the length scale, and hence wavelength, doubles the effect.
The extra term is thus $-2\int\,\dot{\Phi}$.  Integrating from some early time
until the present, one obtains $\Delta T/T=-2\Phi$. 
As an aside, in the metric picture this term arises since the Lagrangian
${\cal L}\propto g_{\mu\nu}\dot{x}^{\mu}\dot{x}^{\nu}$ depends on time.
Since $t$ is not a cyclic coordinate, energy is not conserved along the
photon trajectory.  The redshift goes as $a(t)$ plus the ``extra'' time
dependence from $\Phi(t)$ (e.g.~White et al.~\cite{WSS}).

\section{Conclusions}

We have presented a pedagogical derivation of the coefficient which relates
the large-angle CMB temperature fluctuations to the gravitational potential.
For adiabatic fluctuations, this comes about by a partial cancellation of
two terms -- the intrinsic temperature perturbation and the gravitational
redshift from climbing out of a potential.
The latter wins, meaning photon overdensities are CMB cold spots.
This cancellation, which cannot be present in models with isocurvature initial
conditions, is crucial to the success of the inflationary cold dark matter
model in predicting small CMB fluctuations for a given amount of large-scale
structure (e.g.~White \& Scott \cite{WS}).

\section*{Acknowledgements}

We would like to thank Douglas Scott for useful comments and discussions.
MW thanks John Peacock for a conversation which led us to pursue this
question rigorously.
This work was supported in part by grants from the NSF, the
W.M. Keck Foundation, and the DOE.

\end{document}